# BI-CROSS-VALIDATION OF THE SVD AND THE NONNEGATIVE MATRIX FACTORIZATION[1]

By Art B. Owen and Patrick O. Perry

*Stanford University*

This article presents a form of bi-cross-validation (BCV) for choosing the rank in outer product models, especially the singular value decomposition (SVD) and the nonnegative matrix factorization (NMF). Instead of leaving out a set of rows of the data matrix, we leave out a set of rows and a set of columns, and then predict the left out entries by low rank operations on the retained data. We prove a self-consistency result expressing the prediction error as a residual from a low rank approximation. Random matrix theory and some empirical results suggest that smaller hold-out sets lead to more over-fitting, while larger ones are more prone to under-fitting. In simulated examples we find that a method leaving out half the rows and half the columns performs well.

**1. Introduction.** Many useful methods for handling large data sets begin by approximating a matrix $X \in \mathbb{R}^{m \times n}$ by a product $LR$, where $L \in \mathbb{R}^{m \times k}$ and $R \in \mathbb{R}^{k \times n}$ both have rank $k < \min(m, n)$. Such outer product models are widely used to get simpler representations of large data matrices, either for regularization or for interpretation. The best known example is the truncated singular value decomposition (SVD) [Eckart and Young (1936); Hansen (1987)]. Other examples arise by placing constraints on the factors $L$ and $R$. For instance, the nonnegative matrix factorization (NMF) [see Lee and Seung (1999)] requires $L$ and $R$ to have elements in $[0, \infty)$ and the familiar $k$-means clustering of rows of $X$ imposes a binary structure on $L$. These and some other examples are described in Lazzeroni and Owen (2002).

For all of these methods, the choice of $k$ is problematic. Consider the SVD. The best rank $k$ approximation to $X$, judging by squared error, is obtained by truncating the SVD to $k$ terms yielding $\hat{X}^{(k)}$. Larger values of $k$ provide

Received September 2007; revised December 2008.
[1]Supported by NSF Grant DMS-06-04939.

a better fit to the original $X$ and for $k = \min(m,n)$ we get $\widehat{X}^{(k)} = X$. This clearly raises some risk of overfitting. If $X$ is generated from a noisy process, then that noise is reasonably expected to affect $\widehat{X}^{(k)}$ more strongly when $k$ is large.

We would like to cross-validate the rank selection in order to counter such overfitting. Hoff (2007) states that the usual practice for choosing $k$ is to look for where the last large gap or elbow appears in a plot of singular values. Wold (1978) mentions a strategy of adding terms until the residual standard error matches the noise level, when the latter is known. Cross-validation (CV) is operationally simpler. We do not need to know the noise level, and instead of judging which bend in the curve is the most suitable knee, we select $k$ to minimize a numerical criterion.

Several methods have been proposed for cross-validating the SVD. Of particular interest is the approach taken by Gabriel (2002). Writing $X$ as

$$X = \begin{pmatrix} X_{1,1} & X_{1,2:n} \\ X_{2:m,1} & X_{2:m,2:n} \end{pmatrix},$$

Gabriel fits a truncated SVD of $k$ terms to $X_{2:m,2:n}$, yielding $\widehat{X}^{(k)}_{2:m,2:n}$, and then scores the error in predicting $X_{11}$ by $\widehat{\varepsilon}^{(k)}_{11} = X_{11} - X_{1,2:n}(\widehat{X}^{(k)}_{2:m,2:n})^+ X_{2:m,1}$, where $Z^+$ denotes the Moore–Penrose pseudo-inverse of the matrix $Z$. Leaving out every point in turn, the cross-validated squared error is $\sum_{i=1}^{m}\sum_{j=1}^{n}(\widehat{\varepsilon}^{(k)}_{ij})^2$, where $\widehat{\varepsilon}^{(k)}_{ij}$ is defined analogously to $\widehat{\varepsilon}^{(k)}_{11}$. Gabriel's approach is a kind of bi-cross-validation (BCV), as it leaves out a row and column simultaneously.

Our main goal in this paper is to develop BCV into a tool that is generally applicable to outer product approximations, just as CV is for i.i.d. sampling. We generalize the BCV to $r \times s$ holdouts because in large problems $1 \times 1$ holdouts are unworkable. As a result, there is an $(h \times \ell)$-fold BCV that is a direct analogue of $k$-fold CV. An advantage of CV is that it applies easily for different methods and error criteria. We show here how BCV can be extended to more general outer product models. We use the NMF to show how that works in detail.

The NMF is often thought of as a bi-clustering of rows and columns, and of course $k$-means is clearly a clustering method. The BCV we present thus provides a form of cross-validation for unsupervised learning.

Our second goal is to understand BCV. Gabriel (2002) focuses on the bi-plot and does not explain or motivate BCV. We show a self-consistency property for BCV whereby the residual $\widehat{\varepsilon}^{(k)}$ vanishes for some matrices $X$ of rank $k$. This residual can be explained in terms of principal-components regressions, and it can be generalized in several ways.

The outline of this paper is as follows. Section 2 provides some background on MacDuffee's theorem and the SVD. Section 3 describes how the rank of an SVD is chosen in the crop science literature and explains why the solutions prior to Gabriel (2002) are unsatisfactory. Then it presents the self-consistency lemma mentioned above. Section 4 looks at rank $k$ matrices that fail to satisfy an assumption of the self-consistency lemma. These involve a kind of degeneracy in which the entire signal we seek to detect has been held out. Section 5 shows how to extend BCV from the SVD to the NMF and similar methods. Section 6 applies some recent results from random matrix theory to BCV of the SVD with Gaussian noise. From this analysis, we expect small holdouts to be more prone to overfitting and large holdouts more prone to underfitting. Section 7 has some numerical examples for BCV of the SVD. Section 8 has examples of BCV for the NMF. Section 9 summarizes the results.

1.1. *Related methods.* Many alternatives to cross-validatory choice are based on attempts to count the number of parameters in a rank $k$ model. Some of these methods and their difficulties are mentioned in Section 3.

There are a great many classical methods for choosing $k$ in the context of principal components analysis. There one usually has a number $m \to \infty$ of i.i.d. rows (with variance $\Sigma$) while the number of columns $n$ is fixed. Mardia, Kent and Bibby (1979), Chapter 9.5, presents a likelihood ratio test for whether the last $n - k$ eigenvalues of $\Sigma$ are identical for Gaussian data. Holmes-Junca (1985) compares cross-validation and the bootstrap. Jolliffe (2002), Chapter 6, surveys many methods including techniques based on scree plots, broken stick models and the bootstrap. Scree plots can be quite unreliable. It has long been known that even when all $n$ eigenvalues of $\Sigma$ are identical, there will be some slope in the scree plot, and numerical work in Jackson (1993) finds that a method based on scree plots gives poor performance in some numerical examples motivated by ecology. His simulations also included broken stick models, some bootstraps, some likelihood ratio tests and the Kaiser–Guttman rule that retains any principal component whose eigenvalue is above average. Some broken stick models worked best.

Minka (2000) takes a Bayesian model selection approach. He obtains a posterior distribution on the data matrix under a Gaussian model, with the choice of $k$ corresponding to an assumption that the smallest $n - k$ eigenvalues of $\Sigma$ are equal. The chosen $k$ is the one that maximizes a Laplace approximation to the posterior probability density of the data set. The Laplace method gave the best results in simulations. It beat a cross-validatory method based on holding out rows of the data matrix.

Our setting is different from the ones for principal components. We will have $m$ and $n$ both tending to infinity together and neither rows nor columns need to be i.i.d.

Hoff (2007) treats the SVD under a model where $X$ is a random isotropic low rank matrix plus i.i.d. Gaussian noise. He presents a Gibbs sampling approach for that setting.

The microarray literature has some similar methods for imputing missing values. Troyanskaya et al. (2001) fit an SVD model to the nonmissing values, using an EM style iteration, and then the low rank fitted values are used for imputation. Oba et al. (2003) present a Bayesian version of SVD imputation, using an EM iteration.

The present setting is different from the imputation problems. We have all the values of $X$ and we want to smooth some noise out of it via low rank approximation. The retained data set will consist of rectangular submatrices of $X$ so that the SVD or NMF or other algorithms can be applied to it without requiring missing data methods.

S. Wold (1978) cross-validates the rank of an SVD model by leaving out a scattered set of matrix elements. He advocates splitting the data set into 4 to 7 groups. In his Figure 1, each such group corresponds to one or more pseudo-diagonals of the data matrix, containing elements $X_{ij}$, where $j = i + r$, for some $r$. Having left out scattered data, Wold has to solve an imputation problem in order to define residuals. He does this via the NIPALS algorithm, a form of alternating least squares due to H. Wold (1966).

**2. Background matrix algebra.** This section records some facts about matrix algebra for later use.

The Moore–Penrose inverse of a matrix $X \in \mathbb{R}^{m \times n}$ is denoted by $X^+ \in \mathbb{R}^{n \times m}$. The monograph of Ben-Israel and Greville (2003) provides a detailed presentation of the Moore–Penrose inverse and other generalized inverses. We will also treat scalars as $1 \times 1$ matrices, so that $x^+$ for $x \in \mathbb{R}$ is $1/x$ when $x \neq 0$ and is 0 when $x = 0$.

If matrices $A$ and $B$ are invertible, then $(AB)^{-1} = B^{-1}A^{-1}$ when the products are well defined. Such a reverse-order law does not necessarily hold when the inverse is replaced by a generalized inverse.

There is a special case where a reverse order law holds. Ben-Israel and Greville (2003), Chapter 1.6, credit a private communication of C. C. Mac-Duffee for it. Tian (2004) describes further sufficient conditions for a reverse order law.

THEOREM 1 (MacDuffee's theorem). *Suppose that $X = LR$, where $L \in \mathbb{R}^{m \times k}$ and $R \in \mathbb{R}^{k \times n}$ both have rank $k$. Then*

$$(2.1) \qquad X^+ = R^+ L^+ = R'(RR')^{-1}(L'L)^{-1}L'.$$

PROOF. We find directly from the properties of a Moore–Penrose inverse that $L^+ = (L'L)^{-1}L'$ and $R^+ = R'(RR')^{-1}$. Finally, the RHS of (2.1) satisfies the four defining properties of the Moore–Penrose inverse of $LR$. □

The SVD is discussed in detail by Golub and Van Loan (1996). Let $X \in \mathbb{R}^{m \times n}$ and set $p = \min(m, n)$. The SVD of $X$ is, in it's "skinny" version, $X = U\Sigma V'$, where $U \in \mathbb{R}^{m \times p}$ and $V \in \mathbb{R}^{n \times p}$ with $U'U = V'V = I_p$ and $\Sigma \in \mathbb{R}^{p \times p} = \text{diag}(\sigma_1, \sigma_2, \ldots, \sigma_p)$ with $\sigma_1 \geq \sigma_2 \geq \cdots \geq \sigma_p \geq 0$. The SVD provides an explicit formula $X^+ = V\Sigma^+ U' = \sum_{i=1}^{p} \sigma_i^+ v_i u_i'$ for the Moore–Penrose inverse of $X$.

In applications the SVD is very useful because it can be used to construct an optimal low rank approximation to $X$. For $X \in \mathbb{R}^{m \times n}$, we use $\|X\|_F$ to denote $\sqrt{\sum_{i=1}^{m} \sum_{j=1}^{m} X_{ij}^2}$, its Frobenius norm. The approximation theorem of Eckart and Young (1936) shows that the rank $k$ matrix closest to $X$ in Frobenius norm is

$$\widehat{X}^{(k)} \equiv \sum_{i=1}^{k} \sigma_i u_i v_i'.$$

**3. Cross-validating the SVD.** The idea behind cross-validating the SVD is to hold out some elements of a matrix, fit an SVD to some or all of the rest of that matrix, and then predict the held out portion.

Several strategies have been pursued in the crop science literature. There $i$ typically indexes the genotype of a plant and $j$ the environment in which it was grown, and then $Y_{ij}$ is the yield of food or fiber from plant type $i$ in environment $j$. The value $X_{ij}$ is commonly a residual of $Y_{ij}$ after taking account of some covariates. A low rank approximation to $X_{ij}$ can be used to understand the interactions in yield.

We will ignore the regression models for now and focus on choosing $k$ for the SVD. What makes model selection hard for the SVD is that methods based on counting the number of parameters used in an SVD model do not give reliable $F$ tests for testing whether a setting known to have at least rank $k$ actually has rank $k+1$ or larger. dos S. Dias and Krzanowski (2003) describe some $F$ tests for discerning between ranks $k=0$ and 1 that reject 66% of the time when the nominal rate should be 5%. Some other tests remedy that problem but become too conservative for $k > 0$.

3.1. *Leaving out a matrix element.* Suppose that we wish to hold out the entry $X_{ij}$ and predict it by some $\widehat{X}_{ij}$ computed from the other elements. The best known method is due to Eastment and Krzanowski (1982). They fit an SVD ignoring row $i$ and another one ignoring column $j$. They use the left singular vectors from the SVD that ignored column $j$ and the right singular vectors from the SVD that ignored row $i$. Therefore, $X_{ij}$ is not used in either of those SVDs. From two SVDs they get two sets of singular values. They retain the first $k \leq \min(m-1, n-1)$ of them and combine them via geometric means. Louwerse, Smilde and Kiers (1999) generalize this cross-validation method [as well as that of Wold (1978)] to choose the rank of models fit to $m \times n \times k$ arrays of data.

Unfortunately, separate row and column deletion typically yields cross-validated squared errors that decrease monotonically with $k$ [dos S. Dias and Krzanowski (2003)]. Besse and Ferré (1993) present an explanation of this phenomenon using perturbation theory. As a result, some awkward adjustments based on estimated degrees of freedom are used in practice.

A second difficulty with this kind of cross-validation is that the sign for each term in the combined SVD is not well determined. Eastment and Krzanowski (1982) chose the sign of the $k$th outer product so that its upper left element has the same sign as that from a full SVD fit to the original data. As such, the method does not completely hold out $X_{11}$.

The approach of Gabriel (2002) described in Section 1 does not require looking at $X_{11}$, and its cross-validated squared errors seem to give reasonable nonmonotonic answers in the crop science applications. Accordingly, this is the cross-validation method that we choose to generalize.

3.2. *Leaving out an $r$ by $s$ submatrix.* Suppose that we leave out an $r$ by $s$ submatrix of the $m$ by $n$ matrix $X$. For notational convenience, we suppose that it is the upper left submatrix. Then we partition $X$ as follows

$$(3.1) \qquad X = \begin{pmatrix} A & B \\ C & D \end{pmatrix},$$

where $A \in \mathbb{R}^{r \times s}$, $B \in \mathbb{R}^{r \times (n-s)}$, $C \in \mathbb{R}^{(m-r) \times s}$ and $D \in \mathbb{R}^{(m-r) \times (n-s)}$.

LEMMA 1 (Self consistency). *Suppose that $X \in \mathbb{R}^{m \times n}$ having rank $k$ is partitioned as in equation (3.1) and that the matrix $D \in \mathbb{R}^{(m-r) \times (n-s)}$ appearing there also has rank $k$. Then*

$$(3.2) \qquad A = BD^+C = B(\widehat{D}^{(k)})^+C.$$

PROOF. Because $D$ has rank $k$, we find that $D = \widehat{D}^{(k)}$ and so we only need to prove the first equality. If $k = 0$, then $A = BD^+C = 0$ holds trivially, so we suppose that $k > 0$.

We write the SVD of $X$ as $X = \sum_{i=1}^{k} \sigma_i u_i v_i' = U\Sigma V'$, for $\Sigma = \mathrm{diag}(\sigma_1, \ldots, \sigma_k)$ with $\sigma_1 \geq \sigma_2 \geq \cdots \geq \sigma_k > 0$, $U \in \mathbb{R}^{m \times k}$ and $V \in \mathbb{R}^{n \times k}$. Let $U_1 \in \mathbb{R}^{r \times k}$ contain the first $r$ rows of $U$ and $U_2 \in \mathbb{R}^{(m-r) \times k}$ contain the last $m - r$ rows. Similarly, let $V_1$ contain the first $s$ rows of $V$ and $V_2$ contain the last $n - s$ rows of $V$. Then

$$A = U_1 \Sigma V_1', \qquad B = U_1 \Sigma V_2', \qquad C = U_2 \Sigma V_1' \quad \text{and} \quad D = U_2 \Sigma V_2'.$$

Let $D = LR$, where $L = U_2 S$ and $R = SV_2'$, for $S = \mathrm{diag}(\sqrt{\sigma_1}, \ldots, \sqrt{\sigma_k})$. Then by MacDuffee's theorem,

$$D^+ = R^+ L^+ = R'(RR')^{-1}(L'L)^{-1}L' = V_2 S(SV_2'V_2 S)^{-1}(SU_2'U_2 S)^{-1} SU_2'$$
$$= V_2(V_2'V_2)^{-1}\Sigma^{-1}(U_2'U_2)^{-1}U_2'.$$

Substituting this formula for $D^+$ into $BD^+C$ and simplifying gives

$$\begin{aligned}BD^+C &= (U_1\Sigma V_2')V_2(V_2'V_2)^{-1}\Sigma^{-1}(U_2'U_2)^{-1}U_2'(U_2\Sigma V_1')\\ &= U_1\Sigma V_1' = A. \qquad \square\end{aligned}$$

Lemma 1 would have been quite a bit simpler had we been able to write $D^+$ as $V_2\Sigma^{-1}U_2'$. But $D^+$ need not take that simple form, because the decomposition $D = U_2\Sigma V_2$ is not in general an SVD. The matrices $U_2$ and $V_2$ are not necessarily orthogonal.

We use Lemma 1 to interpret $A - B(\widehat{D}^{(k)})^+C$ as a matrix of residuals for the upper left corner of $X$ with respect to a model in which $X$ follows an SVD of rank $k$. To do $(h \times \ell)$-fold BCV of the SVD, we may partition the rows of $X$ into $h$ subsets, partition the columns of $X$ into $\ell$ subsets, and write

$$\text{BCV}(k) = \sum_{i=1}^{h}\sum_{j=1}^{\ell}\|A(i,j) - B(i,j)(\widehat{D}(i,j)^{(k)})^+C(i,j)\|_F^2,$$

where $A(i,j)$ represents the held out entries in bi-fold $(i,j)$ and similarly for $B(i,j)$, $C(i,j)$ and $D(i,j)$.

There are other ways to define residual quantities. When both $X$ and $D$ have rank $k$, then $A - BD^+C = 0$. We could replace any or all of $A$, $B$, $C$ and $D$ by a rank $k$ approximation and still get a matrix of 0s for $X$ and $D$ of rank $k$. Assuming that we always choose to use $(\widehat{D}^{(k)})^+$, there are still eight different choices for residuals. The simulations in Section 7 consider residuals of the following two types:

(3.3) $\qquad\qquad$ (I) $\quad A - B(\widehat{D}^{(k)})^+C,$

(3.4) $\qquad\qquad$ (II) $\quad A - \widehat{B}^{(k)}(\widehat{D}^{(k)})^+\widehat{C}^{(k)}.$

In the next section we show how residuals of the first type above correspond to a cross-validation of principal components regression (PCR).

3.3. *Cross-validation of regression.* In multivariate multiple regression we have a design matrix $Z \in \mathbb{R}^{m \times n}$ and a response matrix $Y \in \mathbb{R}^{m \times s}$ to be predicted by a regression model of the form $Y \doteq Z\beta$, where $\beta \in \mathbb{R}^{n \times s}$. The case of ordinary multiple regression corresponds to $s = 1$. In either case the least squares coefficient estimates are given by $\widehat{\beta} = (Z'Z)^{-1}Z'Y$, assuming that $Z$ has full rank, and by $\widehat{\beta} = Z^+Y$ more generally.

If we leave out the first $r$ rows of $Z$ and $Y$ and then predict the left out rows of $Y$, we get a residual

$$Y_{1:r,1:s} - Z_{1:r,1:n}\widehat{\beta} = A - BD^+C,$$

in the decomposition

$$(3.5) \qquad (Y \ Z) = \begin{pmatrix} Y_{1:r,1:s} & Z_{1:r,1:n} \\ Y_{(r+1):m,1:s} & Z_{(r+1):m,1:n} \end{pmatrix} \equiv \begin{pmatrix} A & B \\ C & D \end{pmatrix}.$$

From this example, we see that BCV changes regression cross-validation in two ways. It varies the set of columns that are held out, and it changes the rank of the linear model from the number of columns of $Z$ to some other value $k$. The regression setting is much simpler statistically, because then the matrix $D$ to be inverted is not random.

The generalized Gabriel bi-cross-validation we present can now be described in terms of ordinary cross-validation of PCR. Suppose that we hold out the first $r$ cases and fit a regression of $Y_{(r+1):m,1:s}$ on the first $k$ principal components of $Z_{(r+1):m,1:n}$. The regression coefficients we get take the form $(\widehat{D}^{(k)})^+ C$. Then the predictions for the held out entries are $\widehat{A} = B(\widehat{D}^{(k)})^+ C$. Thus, conditionally on the set of columns held out, BCV is doing a CV of PCR.

3.4. *Separate row and column deletion.* The method in Eastment and Krzanowski (1982) does not generally have the self-consistency property. Consider the rank 1 matrix

$$X = \sigma u v' \equiv \sigma \begin{pmatrix} u_0 \\ u_1 \end{pmatrix} \begin{pmatrix} v_0 \\ v_1 \end{pmatrix}',$$

where $\sigma > 0$, $u_0 \in \mathbb{R}$, $u_1 \in \mathbb{R}^{m-1}$, $v_0 \in \mathbb{R}$ and $v_1 \in \mathbb{R}^{n-1}$ with $u'u = v'v = 1$. Leaving out column 1 of $X$ and taking the SVD yields $\sigma(\pm u)(\pm v_1)' = (\sigma \|v_1\|)(\pm u)(\pm v_1/\|v_1\|)'$. The signs in these two factors must match but can be either positive or negative. Similarly, leaving out row 1 of $X$ and taking the SVD yields $(\sigma \|u_1\|)(\pm u_1/\|u_1\|)(\pm v)'$.

Combining these two parts yields

$$\pm \|u_1\|^{-1/2} \|v_1\|^{-1/2} \sigma u v' = \pm \sigma u v' \frac{1}{\sqrt{(1-u_0^2)(1-v_0^2)}}.$$

The sign is arbitrary because the choices from the two SVDs being combined might not match. Even with the correct sign, the estimate is slightly too small in magnitude.

**4. Exceptions.** Lemma 1 has a clause that requires the submatrix $D$ to have rank $k$ just like the full matrix $X$. This clause is necessary. For example, the "spike matrix"

$$(4.1) \qquad X = \begin{pmatrix} 1 & 0 & 0 & 0 \\ 0 & 0 & 0 & 0 \\ 0 & 0 & 0 & 0 \\ 0 & 0 & 0 & 0 \end{pmatrix} = \begin{pmatrix} 1 \\ 0 \\ 0 \\ 0 \end{pmatrix} (1 \ 0 \ 0 \ 0)$$

clearly has rank 1, but any submatrix that excludes the upper left entry has rank 0. Thus, $A \ne BD^+C$ for this matrix whenever $A$ contains the upper left entry. Clearly, $BD^+C$ equals zero for this matrix. The squared error from fitting the true rank $k = 1$ is the same as that from fitting $k = 0$.

One could not reasonably expect to predict the 1 in the upper left corner of the spike matrix in (4.1) from its other entries that are all 0s. In this instance, the exception seems to provide a desirable robustness property.

Another exception arises for a "stripe matrix" such as

$$(4.2) \qquad X = \begin{pmatrix} 1 & 1 & 1 & 1 \\ 0 & 0 & 0 & 0 \\ 0 & 0 & 0 & 0 \\ 0 & 0 & 0 & 0 \end{pmatrix} = \begin{pmatrix} 1 \\ 0 \\ 0 \\ 0 \end{pmatrix} \begin{pmatrix} 1 & 1 & 1 & 1 \end{pmatrix}.$$

In this case, predicting a 1 for the upper left entry from the rest of the data seems slightly more reasonable, because that is what a continuation of the top row would give. Of course, a continuation of the left most column would give a prediction of 0. For this case it is not so clear whether the upper left value should be predicted to be a 0 or a 1 from the other entries. In crop science applications, the row might correspond to a specific genotype or phenotpye and then it would have been caught by an additive or regression part of the model. Put another way, the residual matrices that arise in crop science do not generally contain additive main effects of the type shown above. However, in other applications, this sort of pattern is plausible.

Yet another exception arises for an "arrow matrix" such as

$$(4.3) \qquad X = \begin{pmatrix} 1 & 1 & 1 & 1 \\ 1 & 0 & 0 & 0 \\ 1 & 0 & 0 & 0 \\ 1 & 0 & 0 & 0 \end{pmatrix} = \begin{pmatrix} 1 & 0 \\ 1 & 1 \\ 1 & 1 \\ 1 & 1 \end{pmatrix} \begin{pmatrix} 1 & 1 & 1 & 1 \\ 0 & -1 & -1 & -1 \end{pmatrix}.$$

This matrix has rank 2, but the upper left entry will not be correctly predicted by the formula $BD^+C$. In this case a value of 1 seems like a plausible prediction for the upper left element based on all the others. It fits a multiplicative model for elements of $X$. But it is not the only plausible prediction because an additive model would predict a 2 for the upper left entry.

It is clear that if the formula $BD^+C$ is to match $A$, that the singular value decomposition of $X$ when restricted to the lower right $m - r$ by $n - s$ submatrix must not become degenerate. The columns of $U_2 \in \mathbb{R}^{(m-r) \times k}$ from the proof of Lemma 1 must not be linearly dependent, nor can those of $V_2 \in \mathbb{R}^{(n-s) \times k}$ be linearly dependent.

If we hold out $r$ rows and $s$ columns, then some outer product feature, which affects fewer than $r + 1$ rows or $s + 1$ columns, can similarly be missed. Therefore, features of the low rank approximation to $X$ that are not sufficiently broadly based are not faithfully represented in bi-cross-validation. In the case of spikes and stripes, this seems to be a desirable noise suppression property. If the data set features a large number of very small tight clusters, such as near duplicates of some rows, then bi-cross-validation might underestimate $k$.

This property of bi-cross-validation is not a surprise. A similar phenomenon happens in ordinary cross-validation. If leaving out a subset of the predictors causes the design matrix to become singular, then cross-validating a correct model with no noise will not give a zero residual.

If one is willing to forgo the robustness of ignoring spikes and is concerned about missing large but very sparse components in the SVD, then there is a remedy. Let $O_L$ and $O_R$ be uniform random orthogonal matrices of dimensions $m \times m$ and $n \times n$ respectively. Then $\widetilde{X} = O_L X O'_R = O_L U \Sigma V' O'_R$ has the same singular values as $X$ but has no tendency to concentrate the singular vectors into a small number of rows or columns. The BCV of $\widetilde{X}$ is equally sensitive to outer products $\sigma u v'$ with sparse $u$ and $v$ as with nonsparse $u$ and $v$.

**5. Cross-validating the NMF and other outer product decompositions.** In the nonnegative matrix factorization of Juvela, Lehtinen and Paatero (1994) and Lee and Seung (1999), $X$ is approximated by a product $WH$, where $W \in [0, \infty)^{m \times k}$ and $H \in [0, \infty)^{k \times n}$. That is, the matrices $W$ and $H$ are constrained to have nonnegative entries. Ordinarily, both factors of $X$ have rank $k$. We will suppose that $W$ and $H$ are chosen to minimize $\|X - WH\|_F^2$, although other objective functions are also used and can be bi-cross-validated. The estimated NMF often has many zero entries in it. Then the rows of $H$ can be interpreted as sparsely supported prototypes for those of $X$, yielding a decomposition of these rows into distinct parts [Lee and Seung (1999)].

There are several algorithms for computing the NMF. We use alternating constrained least squares. Given $W$, we choose $H$ to minimize $\|X - WH\|_F^2$ over $[0, \infty)^{k \times n}$, and given $H$, we choose $W$ to minimize $\|X - WH\|_F^2$ over $[0, \infty)^{m \times k}$. Unlike the SVD, there is no certainty of attaining the desired global optimum.

5.1. *Residuals for the NMF.* To bi-cross-validate the NMF, we proceed as for the SVD. Let $X$ be decomposed into parts $A$, $B$, $C$ and $D$ as before, and suppose that $X = WH$, where $W \in [0, \infty)^{m \times k}$ and $H \in [0, \infty)^{k \times n}$ both have rank $k$. Then if the retained submatrix $D$ also has rank $k$, we once again find $A = BD^+C$ by applying the self-consistency Lemma 1, followed by MacDuffee's theorem. Indeed,

$$(5.1) \qquad A = B(\widehat{D}^{(k)})^+ C = B(W_D H_D)^+ C = (BH_D^+)(W_D^+ C),$$

where $D = W_D H_D$ is the NMF for $X$ restricted to the submatrix $D$. This process leads to the residual matrix

(5.2) $$A - \widehat{W}_A^{(k)} \widehat{H}_A^{(k)},$$

where $\widehat{W}_A^{(k)} = B(\widehat{H}_D^{(k)})^+$ and $\widehat{H}_A^{(k)} = (\widehat{W}_D^{(k)})^+ C$, and $D = \widehat{W}_D^{(k)} \widehat{H}_D^{(k)}$ is an NMF fit to $D$. We refer to (5.2) as the "simple" residual below. It vanishes when a $k$-term NMF holds for both $X$ and $D$.

In general, the left and right factors $\widehat{W}_A^{(k)}$ and $\widehat{H}_A^{(k)}$ in (5.2) need not have nonnegative elements, especially when the rank of $X$ is not equal to $k$. It seems preferable, at least aesthetically, to require both factors to have nonnegative entries. There is a natural way to impose constraints on these factors. The product $(\widehat{W}_D^{(k)})^+ C$ has a least squares derivation which we can change to constrained least squares, replacing

$$\widehat{H}_A^{(k)} = \underset{H \in \mathbb{R}^{k \times s}}{\arg\min} \|C - \widehat{W}_D^{(k)} H\|_F^2 \quad \text{by}$$

$$\widetilde{H}_A^{(k)} = \underset{H \in [0,\infty)^{k \times s}}{\arg\min} \|C - \widehat{W}_D^{(k)} H\|_F^2.$$

We similarly replace $\widehat{W}^{(k)}$ by

$$\widetilde{W}_A^{(k)} = \underset{W \in [0,\infty)^{r \times k}}{\arg\min} \|B - W \widehat{H}_D^{(k)}\|_F^2.$$

The residual we use is then

(5.3) $$A - \widetilde{W}_A^{(k)} \widetilde{H}_A^{(k)}.$$

We call (5.3) the "conforming" residual. Unlike simple residuals, the conforming residuals can depend on which factorization of $D$ is used. If $X$ has a $k$ term NMF and the submatrix $D$ has an NMF unique up to permutation and scaling, then the conforming residual is zero. When $D$ has a $k$-term but inherently nonunique NMF, then it may be possible to get a nonzero conforming residual. Laurberg et al. (2008) give necessary and sufficient conditions for the NMF to be unique up to permutation and scaling.

Figure 1 illustrates the BCV process for the NMF. Ordinarily there are $h$ distinct row holdout sets and $\ell$ distinct column sets and then there are $M = h\ell$ holdout operations to evaluate, but the algorithm is more general. For example, one could choose $M$ random $r \times s$ submatrices of $X$ to hold out.

Requiring nonnegative factors in the estimate of $A$ is not analogous to replacing the type I residuals (3.3) by type II residuals (3.4). The analogous modification would be to replace the submatrices $B$ and $C$ by their rank $k$ NMF fits. We do not investigate this option.

The examples from Section 4 have implications for bi-cross-validation of the NMF. If $X = WH = \sum_{i=1}^k w_i h_i'$ and one of the vectors $w_i$ or $h_i$ is almost entirely zeros, then that factor will be harder to resolve.

**Given:**

A matrix $X \in [0, \infty)^{m \times n}$, row and column holdout subsets $\mathcal{I}_\ell \subset \{1, \ldots, m\}$, $\mathcal{J}_\ell \subset \{1, \ldots, n\}$ for $\ell = 1, \ldots, M$, and a list of ranks $\mathcal{K} \subset \{1, \ldots, \min(m, n)\}$.

**Do:**

1) For $k \in \mathcal{K}$: $\text{BCV}(k) \leftarrow 0$
2) For $\ell \in \{1, \ldots, L\}$ and $k \in \mathcal{K}$:
3) $\quad \mathcal{I} \leftarrow \mathcal{I}_\ell$ and $\mathcal{J} \leftarrow \mathcal{J}_\ell$
4) $\quad$ Fit NMF: $X_{-\mathcal{I},-\mathcal{J}} \doteq W^{(k)}_{-\mathcal{I},-\mathcal{J}} H^{(k)}_{-\mathcal{I},-\mathcal{J}}$
5) $\quad W^{(k)}_{\mathcal{I},\mathcal{J}} \leftarrow \text{argmin}_{W \in [0,\infty)^{r \times n-s}} \|X_{\mathcal{I},-\mathcal{J}} - W H^{(k)}_{-\mathcal{I},-\mathcal{J}}\|_F^2$
6) $\quad H^{(k)}_{\mathcal{I},\mathcal{J}} \leftarrow \text{argmin}_{H \in [0,\infty)^{m-r \times s}} \|X_{-\mathcal{I},\mathcal{J}} - W^{(k)}_{-\mathcal{I},-\mathcal{J}} H\|_F^2$
7) $\quad \widehat{X}^{(k)}_{\mathcal{I},\mathcal{J}} \leftarrow W^{(k)}_{\mathcal{I},\mathcal{J}} H^{(k)}_{\mathcal{I},\mathcal{J}}$
8) $\quad \text{BCV}(k) \leftarrow \text{BCV}(k) + \|X_{\mathcal{I},\mathcal{J}} - \widehat{X}^{(k)}_{\mathcal{I},\mathcal{J}}\|_F^2$.

**Return:**

$\text{BCV}(k)$ for $k \in \mathcal{K}$.

FIG. 1. *This algorithm describes bi-cross-validation of the nonnegative matrix factorization. It uses a squared error criterion and conforming residuals (5.3). To use the simple residuals (5.2), replace lines 5 through 7 by* $\widehat{X}^{(k)}_{\mathcal{I},\mathcal{J}} \leftarrow X_{\mathcal{I},-\mathcal{J}}(H^{(k)}_{-\mathcal{I},-\mathcal{J}})^+ (W^{(k)}_{-\mathcal{I},-\mathcal{J}})^+ X_{-\mathcal{I},\mathcal{J}}$.

5.2. *Other outer product models.* Other models of outer product type are commonly used on matrix data. Lee and Seung (1999) point out that the model underlying $k$-means clustering of rows has the outer product form $X \doteq LR$, where $L \in \{0,1\}^{m \times k}$ with each row summing to 1 while $R \in \mathbb{R}^{k \times m}$. It may thus be bi-cross-validated in a manner analogous to the NMF, respecting the binary constraints for $L$. It may seem odd to leave out some variables in a cluster analysis, but Hartigan (1975) has advocated leaving out variables in order to study stability of clustering, so bi-cross-validation of $k$-means may be useful. Clusters that involve fewer rows than the number being held out could be missed.

The semi-discrete decomposition of Kolda and O'Leary (1998) approximates $X$ by a matrix of the form $U\Sigma V'$, where $U \in \{-1,0,1\}^{m \times k}$, $V \in \{-1,0,1\}^{n \times k}$ and $\Sigma$ is a nonnegative $k$ by $k$ diagonal matrix. MacDuffee's theorem applies to this decomposition because we can absorb $\Sigma$ into one of the other factors. Thus, a residual like the one based on (5.2) can be constructed for the SDD.

**6. Random matrix distribution theory.** We suppose here that $X$ is a rank 0 or 1 matrix plus i.i.d. Gaussian noise. While we expect BCV to be useful more generally, this special case has a richly developed (and still growing) theory that we can draw on to investigate BCV theoretically. The Gaussian setting allows special methods to be used. Some of the random matrix theory results have been extended to more general distributions, usually requiring finite fourth moments. See Baik and Silverstein (2004) and Soshnikov (2001) for some generalizations.

Section 6.1 considers the case where the true $k = 0$ and we fit a model of either rank 0 or 1. Results on sample covariances of large random matrices are enough to give insight into the hold-out squared error. The case where the data are generated as rank 1 plus noise is more complicated, because it cannot be handled through the sample covariance matrix of i.i.d. random vectors. Section 6.2 reviews some recent results of Onatski (2007) which apply to this case. Then Section 6.3 applies these results to our partitioned matrix. Section 6.4 organizes the findings in a $2 \times 2$ table with true and fitted ranks both in $\{0, 1\}$.

For simplicity, we look at the differences between $\mathbb{E}(\mathrm{BCV}(k))$ for the correct and incorrect $k$. When these are well separated, then we expect the method to more easily select the correct rank. For this analysis, we neglect the possibility that in some settings using the correct rank may give a greater squared error.

6.1. *Pure noise.* Throughout this section $X_{ij}$ are independent $\mathcal{N}(0,1)$ random variables. There is no loss of generality in taking unit variance. We partition $X$ into $A$, $B$, $C$ and $D$ as before, leaving out the $r \times s$ submatrix $A$ and fitting an SVD of rank $k$ to $D$. The true $k$ is 0 and we will compare fits with $k = 0$ to fits with $k = 1$.

For this pure noise setting we will assume that $m \geq n$. This can be arranged by transposing $X$ and simultaneously interchanging $r$ and $s$.

For $k = 0$, we trivially find that $\widehat{D}^{(0)} = 0$ and so the error from a rank 0 fit is $\widehat{\varepsilon}^{(0)} = A - B(\widehat{D}^{(0)})^+ C = A$. For $k = 1$, the residual is $\widehat{\varepsilon}^{(1)} = A - B(\widehat{D}^{(1)})^+ C$. Because $A$ is independent of $B$, $C$ and $D$, we easily find that

$$\mathbb{E}(\|\widehat{\varepsilon}^{(1)}\|_F^2) = \mathbb{E}(\|A\|_F^2) + \mathbb{E}(\|B(\widehat{D}^{(1)})^+ C\|_F^2) \geq \mathbb{E}(\|A\|_F^2) = \mathbb{E}(\|\widehat{\varepsilon}^{(0)}\|_F^2).$$

The true model has an advantage. Its expected cross-validated error is no larger than that of the model with $k = 1$.

The rank 1 approximation of $X$ is comparatively simple in this case. We may write it as $\widehat{X}^{(1)} = \sigma_1 uv'$. Then, from Muirhead (1982) we know that $u$, $v$ and $\sigma_1$ are independent with $u$ uniformly distributed on the sphere $S^{m-1} = \{x \in \mathbb{R}^m \mid x'x = 1\}$, and $v$ uniformly distributed on $S^{n-1}$.

Let $\sigma_1$ be the largest singular value of $X$. Then $\sigma_1^2$ is the largest eigenvalue of $X'X$. Such extreme eigenvalues have been the subject of much recent work. The largest has approximately a (scaled) Tracy–Widom distribution

$W_1$. Specifically, let

$$\mu_{m,n} = (\sqrt{m-1} + \sqrt{n})^2 \quad \text{and}$$

$$\sigma_{m,n} = (\sqrt{m-1} + \sqrt{n})\left(\frac{1}{\sqrt{m-1}} + \frac{1}{\sqrt{n}}\right)^{1/3}.$$

Then from Theorem 1 of Johnstone (2001),

$$\frac{\sigma_1^2 - \mu_{m,n}}{\sigma_{m,n}} \xrightarrow{d} W_1$$

as $m$ and $n$ go to infinity in such a way that $m/n \to c \geq 1$. For our purposes it will be accurate enough to replace the $(m-1)$'s above by $m$. While $\mu_{m,n}$ grows proportionally to $m$, $\sigma_{m,n}$ grows only like $m^{1/3}$, so $\sigma_1^2$ is relatively close to its mean.

Putting this together, $B(\widehat{D}^{(1)})^+ C = (Bu)\sigma_1^{-1}(v'C)$, where $Bu \in \mathbb{R}^{r \times 1}$ and $v'C \in \mathbb{R}^{1 \times s}$ both have i.i.d. $\mathcal{N}(0,1)$ entries, independent of $\sigma_1$, and so

$$\mathbb{E}(\|B(\widehat{D}^{(1)})^+ C\|_F^2) = rs\mathbb{E}(\sigma_1^{-2}).$$

The random variable $\sigma_1^2$ is stochastically larger than a $\chi^2_{(m)}$ random variable and so $\mathbb{E}(\sigma_1^{-2})$ does exist. Also, $\sigma_{m,n}/\mu_{m,n}$ becomes negligible as $m, n \to \infty$. Therefore, $\mathbb{E}(\sigma_1^{-2}) \approx \mu_{m-r,n-s}^{-1} = (\sqrt{m-r} + \sqrt{n-s})^{-2}$, recalling that the SVD is applied to $D$ which has dimensions $(m-r) \times (n-s)$.

When we partition the rows of $X$ into $m/r$ subsets and the columns into $n/s$ subsets and then average the expected errors, we get

(6.1)
$$\frac{1}{mn}\mathbb{E}(\text{BCV}(0)) = 1 \quad \text{and}$$
$$\frac{1}{mn}\mathbb{E}(\text{BCV}(1)) \approx 1 + (\sqrt{m-r} + \sqrt{n-s})^{-2}.$$

The bi-cross-validated squared error is, on average, slightly larger for $k = 1$ than for $k = 0$. As equation (6.1) shows, the expected difference between BCV(1) and BCV(0) grows if $r/m$ and $s/n$ are taken larger. Thus, we expect larger holdouts to be better for avoiding overfitting.

6.2. *Rank 1 plus noise.* Now we suppose that

(6.2) $$X = \kappa uv' + Z$$

for unit vectors $u \in \mathbb{R}^m$ and $v \in \mathbb{R}^n$, and a constant $\kappa > 0$. Here $\kappa uv'$ is a rank 1 signal obscured by the noise, $Z$. The elements $Z_{ij}$ of $Z$ are i.i.d. $\mathcal{N}(0,1)$ random variables. If we fit a rank 0 model, then the expected mean squared error is $\mathbb{E}(\text{BCV}(0)) = \kappa^2 + mn$. The root mean square of the elements in the signal is $\kappa(mn)^{-1/2}$ and so it is reasonable to consider $\kappa$ increasing with $m$ and $n$.

When exactly one of $u$ and $v$ is fixed and the other is random, then model (6.2) is a special case of the spiked covariance model of Johnstone (2001). The spiked covariance model has either independent rows and correlated columns or vice versa. If columns are independent, then $XX'$ is a spiked covariance, and theoretical results on its first eigenvalue apply directly to the left singular vector of $X$. Conversely, independent rows let us approach the left singular vector of $X$ via $X'X$. But we need both left and right singular vectors of $X$ and we cannot assume that both rows and columns are independent.

Some recent results due to Onatski (2007) allow us to investigate model (6.2) where both $u$ and $v$ are fixed. Either or both could be random as well.

We specialize the results of Onatski (2007) to the case of $k = 1$ deterministic factor with noise level $\sigma = 1$ and deterministic loadings. His deterministic factor $F$ is our vector $v\sqrt{n}$. His loadings $L$ then correspond to our $\kappa u/\sqrt{n}$. Onatski (2007) studies a "weak factor model" whose small loadings are such that $L'L = \kappa^2/n$ approaches a constant $d_1$. The weak factor model has $\kappa$ growing slowly with $n$. Earlier work of Bai (2003) used "strong factors" with much larger loadings, in which $L'L/n$ approaches a constant.

THEOREM 2. *Let $X = \kappa u v' + Z$, where $u \in \mathbb{R}^m$ and $v \in \mathbb{R}^n$ are unit vectors, and $Z_{ij}$ are independent $\mathcal{N}(0,1)$ random variables. Suppose that $m$ and $n$ tend to infinity together with $m - cn = o(m^{-1/2})$ for some $c \in (0, \infty)$. Let $\hat{\ell}_1$ be the largest eigenvalue of $XX'/n$. If $d_1 \equiv \kappa^2/n > \sqrt{c}$, then $\sqrt{n}(\hat{\ell}_1 - m_1) \to \mathcal{N}(0, \Sigma_d)$ in distribution, where*

$$m_1 = \frac{(d_1 + 1)(d_1 + c)}{d_1} \quad and \quad \Sigma_d = \frac{2(d_1^2 - c)}{d_1^2}(2d_1 + c + 1).$$

PROOF. This is from Theorem 5 of Onatski (2007). □

When $d_1 < \sqrt{c}$ there is a negative result that simply says $\hat{\ell}_1$ tends to $(1 + \sqrt{c})^2$ regardless of the actual value of $d_1 = \kappa^2/n$.

The critical threshold is met when $d_1 > \sqrt{c}$. That is, $\kappa^2/n > \sqrt{c} = \sqrt{m/n}$, and so the threshold is met when

(6.3) $$\kappa^2 = \delta\sqrt{mn} \quad \text{for } \delta > 1.$$

The critical value for $\kappa$ is unchanged if we transpose $X$, thereby switching $m$ and $n$. We will consider large but finite $\delta$.

The top singular value of $X$ is $\hat{\sigma}_1$ and $\hat{\sigma}_1^2 = n\hat{\ell}_1$, which grows as $m$ and $n$ grow. We scale it by $\kappa^2$ which grows at the same rate, and find

$$\frac{\mathbb{E}(\hat{\sigma}_1^2)}{\kappa^2} = \frac{n\mathbb{E}(\hat{\ell}_1)}{\kappa^2} \approx \frac{nm_1}{\kappa^2} = \frac{n(d_1+1)(d_1+c)}{d_1 \kappa^2} = \left(1 + \frac{1}{\delta\sqrt{c}}\right)\left(1 + \frac{\sqrt{c}}{\delta}\right),$$

after some simplification.

The first singular value $\hat\sigma_1$ then scales as a multiple of $\kappa$. That multiple does not tend to one in the weak factor limit. But if $\delta$ is large, the bias is small.

The variance of $\hat\sigma_1^2/\kappa^2 = (\sqrt{n}/\kappa^2) \times \sqrt{n}\hat\ell_1$ is

$$\mathbb{V}\left(\frac{\hat\sigma_1^2}{\kappa^2}\right) \approx \frac{n\Sigma_d}{\kappa^4} = \frac{n}{\kappa^4}\frac{2(d_1^2-c)}{d_1^2}(2d_1+c+1)$$

$$= \frac{2}{\sqrt{mn}}\frac{\delta^2-1}{\delta^2}\left(2\delta+\sqrt{c}+\frac{1}{\sqrt{c}}\right),$$

after some simplification. Treating $\hat\sigma_1^2$ as an estimate of $\kappa^2$, we find that the ratio $\hat\sigma_1^2/\kappa^2$ has a standard deviation that is asymptotically negligible compared to its bias.

Onatski (2007) also considers the accuracy with which the vectors $u$ and $v$ in our notation are estimated.

THEOREM 3. *Under the conditions of Theorem 2, let $\hat u$ be the first eigenvector of $XX'/n$ and $\hat v$ be the first eigenvector of $X'X/m$. If $d_1 > \sqrt{c}$, then*

$$\Sigma_U^{-1/2}(\hat u'u - \mu_U) \to \mathcal{N}(0,1) \quad\text{and}\quad \Sigma_V^{-1/2}(\hat v'v - \mu_V) \to \mathcal{N}(0,1),$$

*where*

$$\mu_U = \sqrt{\frac{d_1^2-c}{d_1(d_1+1)}} \quad\text{and}\quad \mu_V = \sqrt{\frac{d_1^2-c}{d_1(d_1+c)}}$$

*and as $d_1$ becomes large,*

$$\max(\Sigma_U, \Sigma_V) = O(d_1^{-2}).$$

PROOF. These results come from Theorems 1 and 2 of Onatski (2007). The values for $\mu_U$ and $\mu_V$ are found by direct substitution. From part c of his Theorem 2 with $(i,j) = (t,s) = (1,1)$ and $\phi_{1111} = 0$, we find

$$\Sigma_V = \frac{cd_1(d_1+1)^2}{2(d_1+c)(d_1^2-c)^2}\left(1 + c\left(\frac{d_1+1}{d_1+c}\right)^2\right) - \frac{((d_1+1)^2 - (1-c))^2 c^2}{2d_1(d_1^2-c)(d_1+c)^3}.$$

By inspecting the powers of $d_1$ inside this expression for $\Sigma_V$, we may verify that $\Sigma_V = O(d_1^{-2})$ as $d_1 \to \infty$. From Onatski's Theorem 1, part c, we similarly find

$$\Sigma_U = \frac{(c^2+d_1^4)(d_1+1)}{2d_1(d_1^2-c)^2} + \frac{d_1(c-1)}{2(d_1^2-c)(d_1+1)} - \frac{[(d_1+1)^2 - (1-c)]^2 d_1}{2(d_1^2-c)(d_1+1)^3}.$$

The expression for $\Sigma_V$ can be rewritten as a ratio of two polynomials. At first they both appear to be of eighth degree. But upon inspection, the coefficients of $d_1^8$ and $d_1^7$ vanish from the numerator and the result is asymptotic to $5d_1^{-2}/2$ as $d_1$ grows. □

Onatski's quantity $d_1$ equals $\kappa^2/n = \delta\sqrt{c}$. It follows that both $\mu_U$ and $\mu_V$ become close to 1 for large $\delta$, as $m$ and $n$ increase. Similarly, $\Sigma_U$ and $\Sigma_V$ have small limits when $\delta$ is large.

6.3. *Partitioning $X$.* We write $X \in \mathbb{R}^{m \times n}$ in the form

$$X = \begin{pmatrix} A & B \\ C & D \end{pmatrix} = \kappa \begin{pmatrix} u_1 \\ u_2 \end{pmatrix} \begin{pmatrix} v_1 \\ v_2 \end{pmatrix}' + \begin{pmatrix} Z_{11} & Z_{12} \\ Z_{21} & Z_{22} \end{pmatrix},$$

where $A \in \mathbb{R}^{r \times s}$ and the other parts of $X$ of conforming sizes. The entries of $Z$ are independent Gaussian random variables with mean 0 and variance 1. We will compare rank 0 and 1 predictions of $A$, denoted by $\widehat{A}^{(0)}$ and $\widehat{A}^{(1)}$ respectively.

These estimates take the form $Bg(D)C$, where $g(D) \in \mathbb{R}^{(n-s) \times (m-r)}$ may be thought of as a regularized generalized inverse. For $\widehat{A}^{(0)}$ we take $g(D) = 0$, while for $\widehat{A}^{(1)}$ we take $g(D) = (\widehat{D}^{(1)})^+ = \hat{\kappa}^+ \hat{u}_2 \hat{v}_2'$.

The expected mean square error from the rank 0 model is

$$\mathbb{E}(\|A\|_F^2 \mid u, v, Z_{22}) = \mathbb{E}(\|A\|_F^2 \mid u, v) = rs + \kappa^2 |u_1|^2 |v_1|^2.$$

Summing this MSE over $m/r$ hold out row sets and $n/s$ hold out columns sets, we get $\mathbb{E}(\mathrm{BCV}(0)) = mn + \kappa^2$. For weak factors, the expected cross-validated error of the null model is $mn + \delta\sqrt{mn}$.

For the rank 1 model we build some machinery. Lemma 2 below integrates out the contributions from $Z_{11}$, $Z_{12}$ and $Z_{21}$ for $u$, $v$ and $Z_{22}$ fixed by sampling or by conditioning. Lemma 2 also applies to functions $g$ corresponding to SVDs of rank higher than 1 and to the nonnegative matrix factorization, among other methods.

LEMMA 2. *Let $X$ be as partitioned above with matrix $Z$ having independent entries with $\mathbb{E}(Z_{ij}) = 0$ and $\mathbb{V}(Z_{ij}) = 1$. Let $g(\cdot)$ be a function from $\mathbb{R}^{(m-r) \times (n-s)}$ to $\mathbb{R}^{(n-s) \times (m-r)}$ and let $G = g(D)$. Then*

$$\begin{aligned}\mathbb{E}(\|A - BGC\|_F^2 \mid Z_{22}, u, v) \\ = rs + \kappa^2 \|u_1 v_1' - \kappa u_1 v_2' G u_2 v_1'\|_F^2 + s\kappa^2 \|u_1 v_2' G\|_F^2 \\ + r\kappa^2 \|Gu_2 v_1'\|_F^2 + rs\|G\|_F^2 \\ \leq rs + \kappa^2 (1 - \kappa v_2' G u_2)^2 |u_1|^2 |v_1|^2 \\ + \|G\|_F^2 (\kappa^2 s |u_1|^2 |v_2|^2 + \kappa^2 r |u_2|^2 |v_1|^2 + rs).\end{aligned}$$

PROOF. Decompose

$$\begin{aligned}A - BGC &= \kappa u_1 v_1' + Z_{11} - (\kappa u_1 v_2' + Z_{12})G(\kappa u_2 v_1' + Z_{21}) \\ &= Z_{11} - \kappa u_1 v_2' G Z_{21} - \kappa Z_{12} G u_2 v_1' - Z_{12} G Z_{21} \\ &\quad + \kappa u_1 v_1' - \kappa^2 u_1 v_2' G u_2 v_1'\end{aligned}$$

and sum the squares of the terms to get the equality. Then apply bounds $|Xy| \le \|X\|_F |y|$ for matrix $X$ and vector $y$. □

Next we look at $D = \kappa u_2 v_2' + Z_{22}$ and apply the results from Onatski (2007) in Section 6.2. Let the first term in the SVD of $D$ be $\hat\kappa \hat u_2 \hat v_2'$. Then for a rank 1 fit $g(D) = \hat\kappa^{-1} \hat v_2 \hat u_2'$, and so $\|G\|_F^2 = \hat\kappa^{-2}$ and $\kappa v_2' G u_2 = (\kappa/\hat\kappa) v_2' \hat v_2 \hat u_2' u_2$.

From here we proceed informally in averaging out $Z_{22}$.

From Theorem 2 we have $\hat\kappa^2 \ge \kappa^2$ with overwhelming probability because the bias dominates the variance. Therefore, we will suppose that $\mathbb{E}(\|G\|_F^2 \kappa^2) = \mathbb{E}((\kappa/\hat\kappa)^2) \le 1$. Then

$$\begin{aligned}\mathbb{E}(\|A - BGC\|_F^2) &\le rs + \kappa^2 \mathbb{E}_{22}((1 - \kappa v_2' G u_2)^2) |u_1|^2 |v_1|^2 \\ &\quad + s|u_1|^2 |v_2|^2 + r|u_2|^2 |v_1|^2 + rs\kappa^{-2},\end{aligned}$$

where $\mathbb{E}_{22}$ denotes expectation over $Z_{22}$ with $u$ and $v$ fixed. If we sum over $m/r$ disjoint row blocks, and $n/s$ disjoint columns blocks, then

$$\mathbb{E}(\mathrm{BCV}(1)) \le mn + \kappa^2 \sum_{u_1} \sum_{v_1} |u_1|^2 |v_1|^2 \mathbb{E}_{22}((1 - \kappa v_2' G u_2)^2) + \frac{sn}{r} + \frac{rm}{s} + \frac{mn}{\kappa^2},$$

where the summation is over all $(mn)/(rs)$ ways of picking the rows and columns to hold out as $A$. Because $u$ and $v$ are unit vectors, we always have $|u_2|, |v_2| \le 1$.

Next we turn to $(1 - \kappa v_2' G u_2)^2$. Define unit vectors $\widetilde u_2 = u_2/|u_2|$, and $\widetilde v_2 = v_2/|v_2|$, and let $\widetilde\kappa = \kappa |u_2||v_2|$. Then $D = \kappa u_2 v_2' + Z_{22} = \widetilde\kappa \widetilde u_2 \widetilde v_2' + Z_{22}$. The matrix $D$ now plays the role of $X$ in Theorems 2 and 3. Let $D$ have the largest singular value $\hat\kappa$ with corresponding singular vectors $\hat u_2$ and $\hat v_2$, so that $G = \hat\kappa^{-1} \hat v_2 \hat u_2'$. Then

(6.4) $$\kappa v_2' G u_2 = \frac{\kappa}{\hat\kappa} v_2' \hat v_2 \hat u_2' u_2 = \frac{1}{|u_2||v_2|} \frac{\widetilde\kappa}{\hat\kappa} v_2' \hat v_2 u_2' \hat u_2.$$

Equation (6.4) shows where problems arise if we have held out most of the rows dominating $u$. Then $|u_1|$ is large and $|u_2| = \sqrt{1 - |u_1|^2}$ is small, and of course a similar problem arises when $v_1$ has most of the structure from $v$.

From Theorems 2 and 3 we know that $(\widetilde\kappa/\hat\kappa) v_2' \hat v_2 u_2' \hat u_2$ will be quite close to 1, for large $\delta$. We do not make a delicate analysis of this because we want to consider settings where $|u_2|$ and $|v_2|$ need not be close to 1. We suppose only that $|u_2|^2 \ge \eta$ and $|v_2|^2 \ge \eta$ for some $\eta > 1/2$. For example, if $\eta = 3/4$, then we never hold out more than $1/4$ of the squared signal in $u$ or $v$. This is a very conservative condition when $r$ and $s$ are small such as $r = s = 1$. Now our conservative estimate for $\mathbb{E}_{22}((1 - \kappa v_2' G u_2)^2)$ is simply $(1 - 1/\eta)^2$.

Under this very conservative assumption, we get

$$\mathbb{E}(\|A - BGC\|_F^2) \leq mn + \kappa^2(1 - 1/\eta)^2 + \frac{sn}{r} + \frac{rm}{s} + \frac{mn}{\kappa^2}$$
(6.5)
$$= mn + \delta\sqrt{mn}(1 - 1/\eta)^2 + \frac{sn}{r} + \frac{rm}{s} + \frac{\sqrt{mn}}{\delta}.$$

To see why the assumption is conservative, consider a model in which the components of $u$ and $v$ were generated by sampling $\mathcal{N}(0,1)$ entries which are then normalized and fixed by conditioning. Suppose further that $r$ and $s$ are small. Then $|u_1|$ is about $\sqrt{1 - r/m}$. Similarly, $|v_2|$ is about $\sqrt{1 - s/n}$ and in combination with large $\delta$, we find the coefficient $\mathbb{E}_{22}((1 - \kappa v_2' G u_2)^2)$ of $\delta\sqrt{mn}$ in (6.5) nearly vanishes.

6.4. *Summary.* We summarize the results of this section in a $2 \times 2$ layout given in Table 1. For simplicity, we consider proportional holdouts with $r/m = s/n = \theta$. There is an unavoidable contribution of $mn$ to the squared errors coming from the noise $Z_{11}$ in the held out matrix $A$. If the true $k = 0$, and we fit with $k = 0$ then that is all the error there is. Accordingly, we place a 0 in the upper left entry of the table.

If the true $k = 0$ and we fit with $k = 1$, we get a squared extra error of the form

$$\frac{mn}{(\sqrt{m-r} + \sqrt{n-s})^2} = \frac{1}{1-\theta}\frac{\sqrt{mn}}{\sqrt{c} + 1/\sqrt{c} + 2}.$$

This provides the lower left entry in Table 1. Larger holdouts, as measured by $\theta$, increase the separation between rank 0 and rank 1, when the true rank is 0. Similarly, extreme aspect ratios, where $c$ is far from 1, decrease this separation, so we expect those aspect ratios will lead to more errors in the estimated rank.

TABLE 1
*This table summarizes expected cross-validated squared errors from the text. The lower right entry is conservative as described in the text. The value $\eta \in (1/2, 1)$ represents a lower bound on the proportion not held out, for each singular vector $u$ and $v$. The value $\theta$ is an assumed common value for $r/m$ and $s/n$ and $\delta > 1$ is a measure of signal strength*

| $\mathbb{E}(\mathbf{BCV}(k)) - mn$ | True $k = 0$ | True $k = 1$ |
|---|---|---|
| Fitted $k = 0$ | 0 | $\delta\sqrt{mn}$ |
| Fitted $k = 1$ | $\frac{1}{1-\theta}\frac{\sqrt{mn}}{\sqrt{c}+1/\sqrt{c}+2}$ | $\sqrt{mn}(\delta(1-1/\eta)^2 + c^{3/2} + c^{-3/2} + 1/\delta)$ |

Now suppose that the true model has rank $k = 1$ with a value of $\kappa$ satisfying $\kappa^2 = \delta\sqrt{mn}$ for $\delta > 1$. If we fit a rank 0 model, we get an expected summed squared error of $mn + \delta\sqrt{mn}$. This provides the upper right entry of Table 1.

Finally consider the case where we fit a rank 1 model to a rank 1 dataset. For simplicity, suppose that we never hold out most of $u$ or $v$. Then the retained portion of $u$ has $u_2' u_2 \geq \eta > 1/2$ and, similarly, $v_2' v_2 \geq \eta > 1/2$. For $r/m = s/n = \theta$, equation (6.5) simplifies to

$$mn + \delta\sqrt{mn}(1 - 1/\eta)^2 + \sqrt{mn}(c^{3/2} + c^{-3/2}) + \frac{\sqrt{mn}}{\delta},$$

which gives us the lower right entry in Table 1.

We want the lower right entry of Table 1 to be small to raise the chance that a rank 1 model will be preferred to rank 0. Once again, large aspect ratios, $c \gg 1$, are disadvantageous. The holdout fraction $\theta$ does not appear explicitly in the table. But a large holdout will tend to require a smaller value for $\eta$ which will then raise our bound BCV(1).

In summary, large holdouts make it easier to avoid overfitting but make it harder to get the rank high enough, and extreme aspect ratios make both problems more difficult.

**7. Examples for the truncated SVD.** In this section we consider some simulated examples where $X = \mu + Z$ for which $Z \sim \mathcal{N}(0, I_m \otimes I_n)$ is a matrix of i.i.d. Gaussian noise and $\mu \in \mathbb{R}^{m \times n}$ is a known matrix that we think of as the signal. The best value of $k$ is $k^{\text{opt}} = \arg\min_k \|\widehat{X}^{(k)} - \mu\|^2$, and we can determine this in every sampled realization. For a data determined rank $\widehat{k}$, we can compare $\|\widehat{X}^{(\widehat{k})}\|$ to $\|\widehat{X}^{(k^{\text{opt}})}\|$.

We generate the matrix $\mu$ to have pre-specified singular values $\tau_1 \geq \tau_2 \geq \cdots \geq \tau_{\min(m,n)} \geq 0$ as follows. We sample $Y \sim \mathcal{N}(0, I_m \otimes I_n)$, fit an SVD $Y = U\Sigma V'$ and then take $\mu = UTV'$, where $T = \text{diag}(\tau)$.

We look at two patterns for the singular values. In the binary pattern, $\tau \propto (1, 1, \ldots, 1, 0, 0, \ldots, 0)$ including $k$ nonzero singular values. In the geometric pattern, $\tau \propto (1, 1/2, 1/4, \ldots, 2^{-\min(m,n)})$. Here $\mu$ has full rank, but we expect that smaller $k$ will lead to better recovery of $\mu$. We make sure that the smallest singular values of $\mu$ are quite small so that our identification of $\mu$ as the signal is well behaved.

7.1. *Small simulated example.* We first took $m = 50$ and $n = 40$. The singular values were scaled so that $\|\mu\|^2 = \mathbb{E}(\|Z\|^2)$, making the signal equal in magnitude to the expected noise. For this small example, it is feasible to do a 1 by 1 holdout. Gabriel's method requires 2000 SVDs and the Eastment–Krzanowski method requires 4001 SVDs. The results from 10 simulations

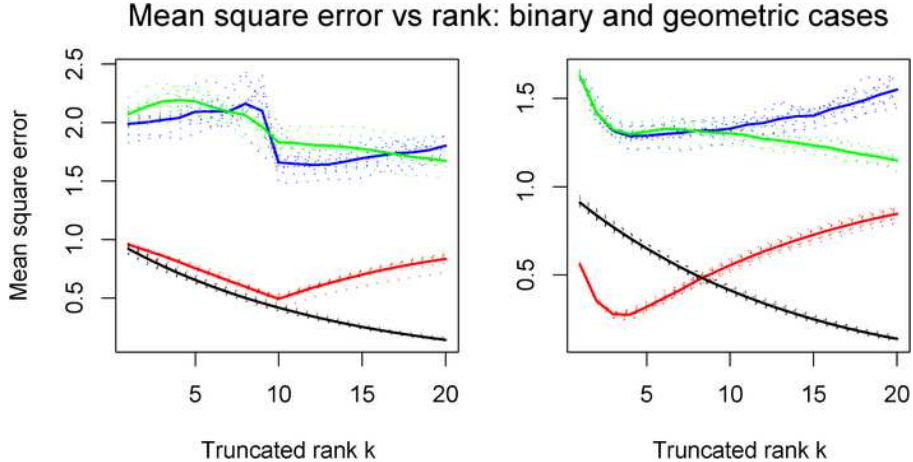

FIG. 2. *This figure shows the results of Gabriel and Eastment–Krzanowski $1 \times 1$ hold out cross-validation on some 50 by 40 matrix examples described in the text. In the left panel the signal matrix has 10 positive and equal singular values and 30 zero singular values. The dotted red curves show the true mean square error per matrix element $\|\widehat{X}^{(k)} - \mu\|^2/(mn)$ from 10 realizations. The solid red curve is their average. Similarly, the black curves show the naive error $\|\widehat{X}^{(k)} - X\|^2/(mn)$. The green curves show the results from Eastment–Krzanowski style cross-validation. The blue curves show Gabriel style cross-validation. The right panel shows a similar simulation for singular values that decay geometrically. In both cases the mean square signal was equal to the expected mean square noise.*

are shown in Figure 2. For the binary pattern, an oracle would have always picked the true rank $k = 10$. For the geometric pattern, the best rank was always 3 or 4. The Eastment–Krzanowski method consistently picked a rank near 20, the largest rank investigated. The Gabriel method tends to pick a rank slightly larger than the oracle would, but not as large as 20.

We also investigated our generalizations of Gabriel's method from Section 3.2 for holdouts of shape $2 \times 2$, $5 \times 5$, $10 \times 10$ and $25 \times 20$. These have residuals given by equations (3.3) and (3.4). For an $r \times s$ holdout, the rows were randomly grouped into $r$ subsets of $m/r$ rows each and the columns were grouped into $s$ subsets of $n/s$ columns each.

For comparison, we also generalized the Eastment–Krzanowski method to $r$ by $s$ holdouts. As for $1 \times 1$ holdouts, one takes the right singular vectors of an SVD on $X$ less $r$ of its rows, the left singular vectors of an SVD on $X$ less $s$ of its columns and assembles them with the geometric means of the two sets of singular vectors. The only complication is in choosing whether to reverse any of the $k$ signs for the singular vector pairs. Let $M_k = u_k v_k'$ be the $k$th term of the SVD fit to $X$ and let $\widehat{M}_k = \hat{u}_k \hat{v}_k'$ be the estimate formed from the two SVDs where part of $X$ was held out. Let $M_k \odot \widehat{M}_k$ be the componentwise product of these matrices. If the mean of $M_k \odot \widehat{M}_k$ over the $rs$ held out elements was negative, then we replaced $\hat{u}_k\hat{v}_k$ by $-\hat{u}_k\hat{v}_k$.

A $2 \times 2$ holdout requires roughly one fourth the work of a $1 \times 1$ holdout. For comparison we also looked at replicating the $2 \times 2$ holdout four times. Each replicate had a different random grouping of rows and columns. We also replicated the $5 \times 5$, $10 \times 10$ and $25 \times 20$ methods 25 times each.

A final comparison was based on three methods from Bai and Ng (2002) that resemble the Bayesian Information Criterion (BIC) of Schwarz (1978). In these, the estimate $\hat{k}$ is the minimizer of

$$(7.1) \quad \mathrm{BIC}_1(k) = \log(\|\widehat{X}^{(k)} - X\|^2) + k\frac{m+n}{mn}\log\frac{mn}{m+n},$$

$$(7.2) \quad \mathrm{BIC}_2(k) = \log(\|\widehat{X}^{(k)} - X\|^2) + k\frac{m+n}{mn}\log C^2 \quad \text{or}$$

$$(7.3) \quad \mathrm{BIC}_3(k) = \log(\|\widehat{X}^{(k)} - X\|^2) + k\frac{\log C^2}{C^2},$$

over $k$, where $c = c(m,n) = \min(\sqrt{m}, \sqrt{n})$.

The methods of Bai and Ng (2002) are designed for a setting where there is a true rank $k < \min(m,n)$ to be estimated. The binary pattern conforms to these expectations but the geometric one does not. In applications, one can seldom be sure whether the underlying pattern has a finite rank, so it is worth investigating (7.1) through (7.3) for both settings.

Figure 3 summarizes all of the methods run for this example. BIC methods 1 and 3 always chose rank $k = 20$, the largest one considered. BIC method 2 always chose rank 1 for the binary setting, but did much better on the geometric setting, for which it is not designed. This example is probably too small for the asymptotics underlying the BIC examples to have taken hold.

The original Eastment–Krzanowski method picked ranks near 20 almost all the time. The generalizations picked ranks near 1 for the binary case but did better on the geometric case.

The lower left region of Figure 3 holds the methods with good relative performance in both cases. The methods there are all generalized Gabriel holdouts for $r \times r$ holdouts with $r \in \{1, 2, 5, 10\}$. These methods have nearly equivalent performance. Both types I and II residuals are there but for no holdout size did type II beat type I, while there were a few cases where the sample mean square was smaller for type I. The methods using replication got slightly better performance on the binary case but somewhat worse performance on the geometric case. There is a slight tendency for larger holdouts to do better on the geometric case, and for smaller ones to do better on the binary case. The pattern we see is reasonable when we notice that the holdout sizes that did poorly for the binary case were not small compared to the number of nonzero singular values.

## Squared error per element

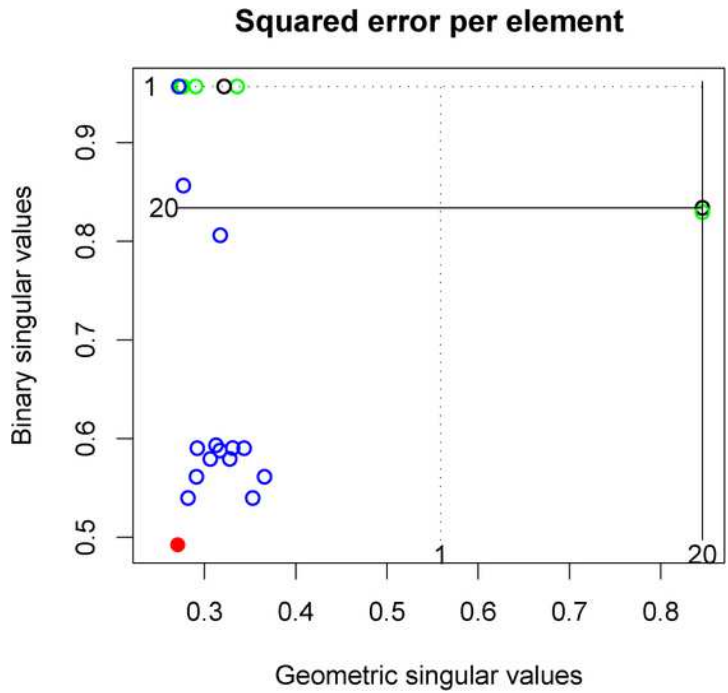

Fig. 3. *This figure shows the mean square error, per element, for all the methods applied to the $50 \times 40$ example. The case with geometrically decaying singular values is on the horizontal axis, and the binary case with ten equal nonzero singular values is on the vertical axis. Gabriel's method and our generalizations are shown in blue, (generalized) Eastment–Krzanowski is in green, the oracle is red, and Bai and Ng's BIC estimators are in black. There are horizontal and vertical reference lines for methods that always pick $k = 1$ or $k = 20$. The cluster of blue points in the lower left corner is discussed in the text.*

The comparisons between I and II at fixed holdout size may not be statistically significant. While one could possibly find significance by pooling carefully over holdout sizes, the more important consideration is that type II BCV is more awkward to implement and appears to perform worse. So these results favor type I, as do further results in the next section. The choice between Gabriel and EK style cross-validation is important. Similarly, the choice of holdout size is important because the largest holdout size did not end up among the most competitive methods.

7.2. *Larger simulated example.* We repeated the simulation described above for $X \in \mathbb{R}^{1000 \times 1000}$. We did not repeat the Eastment–Krzanowski methods because the sign selection process is awkward and the methods are not competitive. We did repeat the Bai and Ng methods because they are easy to implement, and are computationally attractive in high dimensions. We considered both residuals I and II.

For the binary example we used a matrix $\mu$ of rank 50. The geometric example was as before. We found that taking $\|\mu\|^2 = \mathbb{E}(\|Z\|^2)$ makes for a very strong signal when there are $10^6$ matrix elements. So we also considered a low signal setting with $\|\mu\|^2 = 0.01\mathbb{E}(\|Z\|^2)$ and a medium signal setting with $\|\mu\|^2 = 0.1\mathbb{E}(\|Z\|^2)$. The quantity $\delta$ takes values 20, 2 and 0.2 in the high, medium and low signal settings. The low signal setting has $\delta < 1$ and so we expect that not even the largest singular vector should be properly identified.

For the generalized Gabriel BCV we took holdouts of $200 \times 200$, $500 \times 500$ and $800 \times 800$. We used a single random grouping for each holdout.

We looked at approximations of rank $k$ from 0 to 100 inclusive. This requires fitting a truncated SVD of an $m-r$ by $n-s$ matrix to $k = 100$ terms. When $k \ll \min(m-r, n-s)$ the truncated SVD is much faster than the full one. We noticed a large difference using svds in Matlab, but no speed-up using svd in R. In some simulated examples comparable to the present setting, an SVD of $X \in \mathbb{R}^{m \times n}$ to $k$ terms takes roughly $O(mnk)$ computation, at least for large $m \leq n$. The exact cost depends on how well separated the singular values are, and is not available in closed form. A sharper empirical estimate, not needed for our application, could use different powers for $m$ and $n$.

For large $k$ the approximation $(\widehat{D}^{k+1})^+$ can be computed by updating $(\widehat{D}^k)^+$ with an outer product from the SVD of $D$. For type II residuals similar updates can be used for the pieces $B$ and $C$ of $X$. Efficient updating of type II residuals is more cumbersome than for type I.

The outcomes for type I residuals in the two high signal cases are shown in Figure 4. The BCV errors for $800 \times 800$ holdouts were much larger than those for $500 \times 500$ holdouts, which in turn are larger than those for $200 \times 200$ holdouts. The selected rank $\hat{k}$ only depends on relative comparisons for a given holdout, so we have plotted $\mathrm{BCV}(k;r,s)/\mathrm{BCV}(0;r,s)$ versus $k$, in order to compare the minima. A striking feature of the plot is that all 10 random realizations were virtually identical. The process being simulated is very stable for such large matrices.

Comparing the BCV curves in Figure 4, the one for large holdouts is the steepest one at the right end of each panel. Large holdouts do better here at detecting unnecessarily high rank. For the binary case, the steepest curves just left of optimal rank 50 come from small holdouts. They are best at avoiding too small a rank choice.

The type II BCV did not lead to very good results. The only time it was better than type I was for $200 \times 200$ holdouts. There type II matched the oracle all 10 times for the low geometric signal. They were slightly better than type I BCV 9 times out of 10, for the binary medium signal case.

The ratio $\|\widehat{X}^{(\hat{k})} - \mu\|^2 / \|\widehat{X}^{(k^{\mathrm{opt}})} - \mu\|^2$ (or its logarithm) measures the regret we might have in choosing the rank $\hat{k}$ produced by a method. Figure 5

shows these quantities for 6 of the methods studied here. It is noteworthy that BCV with $500 \times 500$ holdouts attained the minimum in all 60 simulated cases. Smaller holdouts had difficulty with medium strength binary singular values and larger holdouts had difficulty with medium strength geometric singular values.

In the binary high signal case the optimal rank, as measured by $\|\widehat{X}^{(k)} - \mu\|^2$, was $k^{\mathrm{opt}} = 50$ in all 10 realizations. The other 5 settings also yielded the same optimal rank for all 10 of their realizations. Those ranks and their estimates are summarized in Table 2. When small holdouts go wrong, they choose $\hat{k}$ too large and conversely for large holdouts, with a slight exception in the first row.

The $200 \times 200$ holdouts were the only method to get close to the true rank 50, in the medium signal binary case. Despite getting nearest the true rank, they had the worst squared errors of the six methods shown. Ironically the BIC methods are designed to estimate rank, but always chose $\hat{k} = 0$ for this setting and did better for it.

7.3. *Real data examples.* The truncated SVD has been recommended for analysis of microarray data by Alter, Brown and Botstein (2000). We have applied BCV to some microarray data reported in Rodwell et al. (2004).

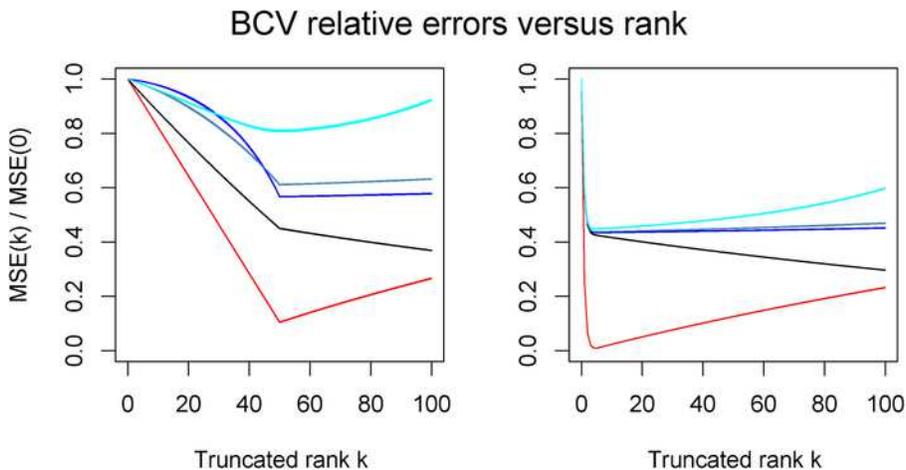

Fig. 4. *This figure shows the BCV errors for the $1000 \times 1000$ examples with equal signal and noise magnitude, as described in the text. The left panel shows the results when there are 50 positive and equal singular values. The horizontal axis is fitted rank, ranging from 0 to 100. The vertical axis depicts square errors, in each case relative to the squared error for rank $k = 0$. There are 10 nearly identical red curves showing the true error in 10 realizations. The black curves show the naive error. Blue curves from dark to light show BCV holdouts of $200 \times 200$, $500 \times 500$ and $800 \times 800$. The right panel shows the results for a geometric pattern in the singular values.*

Squared error relative to oracle

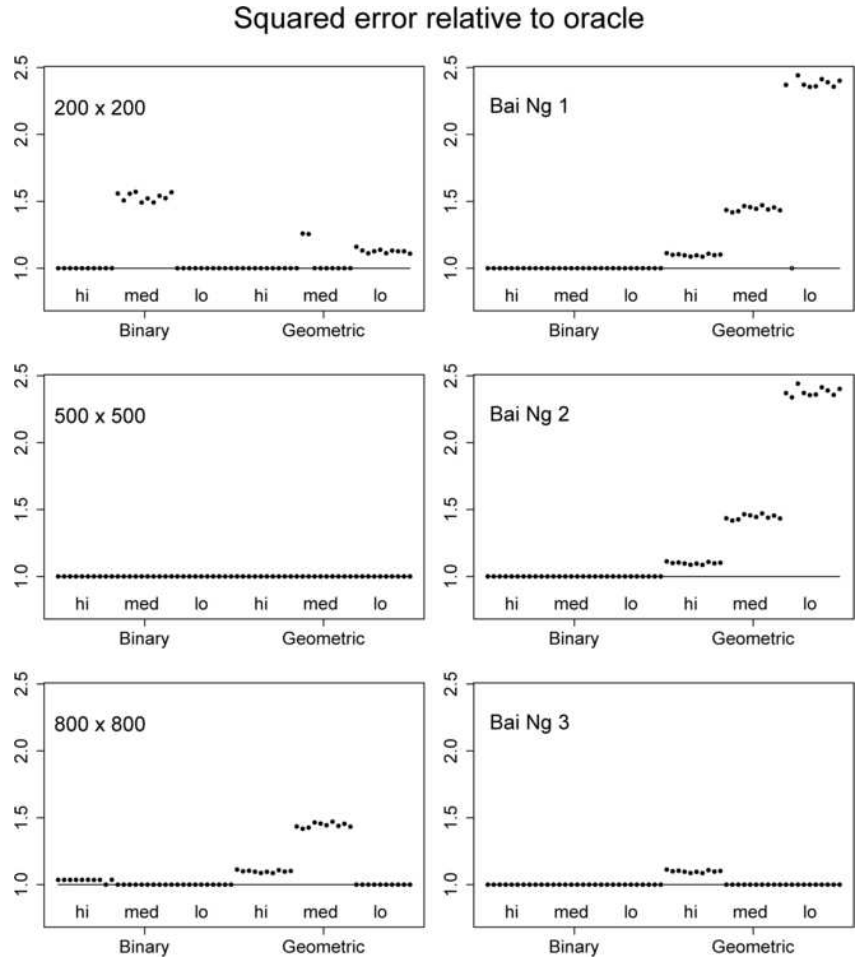

FIG. 5. *This figure shows squared errors attained by rank selection methods relative to an oracle that knows the optimal rank in each simulated $1000 \times 1000$ data set. For each method there are 60 relative errors corresponding to binary vs geometric singular values, high, medium and low signal strength, each replicated 10 times. The methods shown are BCV with $h \times h$ holdouts for $h \in \{200, 500, 800\}$ and the three BIC methods of Bai and Ng (2002).*

The data comprise a matrix of 133 microarrays on 44,928 probes, from a study of the effects on gene expression of aging in the human kidney. The 133 arrays vary with respect to age and sex as well as the tissue type, some being from the kidney cortex while others are from the medulla.

Figure 6 shows the results from bi-cross-validation of the SVD on this data using $(2 \times 2)$-fold holdouts. The rank 0 case has such a large error as to be uninteresting and so the errors are shown relative to the rank 1 case.

TABLE 2
*This table shows the average rank $\hat{k}$ chosen in 10 replications of the $1000 \times 1000$ matrix simulations described in the text. There is one row for each singular value pattern. The column "Opt" shows $\arg\min_k \|\widehat{X}^{(k)} - \mu\|^2$. The next three columns are for $\hat{k}$ chosen via $h \times h$ holdouts for $h \in \{200, 500, 800\}$. The final three columns are for 3 BIC style methods of Bai and Ng (2002). The integer values arose when the same rank was chosen 10 times out of 10*

|  |  | Opt | 200 | 500 | 800 | $BIC_1$ | $BIC_2$ | $BIC_3$ |
|---|---|---|---|---|---|---|---|---|
| Binary | High | 50 | 50 | 50 | 50.9 | 50 | 50 | 50 |
|  | Medium | 0 | 42.6 | 0 | 0 | 0 | 0 | 0 |
|  | Low | 0 | 0 | 0 | 0 | 0 | 0 | 0 |
| Geometric | High | 5 | 5 | 5 | 4 | 4 | 4 | 4 |
|  | Medium | 3 | 3.2 | 3 | 2 | 2 | 2 | 3 |
|  | Low | 1 | 2 | 1 | 1 | 0 | 0 | 1 |

The naive MSE keeps dropping while the BCV version decreases quickly for small $k$ before becoming flat. The minimum BCV error over $1 \leq k \leq 60$ takes place at rank $k = 41$.

We have seen similar very flat BCV curves in other microarray data sets. Microarray data sets tend to have quite extreme aspect ratios. Here the aspect ratio $c = m/n$ is over 300. Simulations with Gaussian data, using binary and geometric singular values and more extreme aspect ratios, behave like the Gaussian data simulations reported earlier in this section, and not like microarray BCV results.

It is plausible that the biological processes generating the microarray data sets contain a large number of features compared to the small number of ar-rays sampled. The singular vectors in the real microarray data examples tend to have high kurtosis, with values in the tens or hundreds. For the kidney data the 133 singular vectors of length 44,928 had an average kurtosis of over 180. Such high kurtosis could arise from small intense clusters corresponding to features, or from heavy tailed noise. But heavy tailed noise values would not be predictive of each other in a holdout, and so we infer that multiple sources of structure are likely to be present.

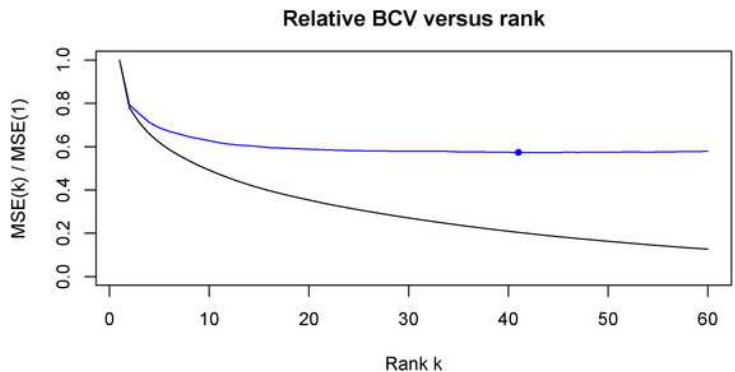

FIG. 6. *This figure compares the naive squared error in black to the $(2 \times 2)$-fold bi-cross–validation, shown in blue. The ranks investigated vary from 1 to 60. Both error curves are normalized by their value at rank $k = 1$. The BCV curve takes it's minimum at $k = 41$, which is marked with a reference point.*

We also ran BCV to estimate the rank of a truncated SVD for some standardized testing data given to us by Ed Haertel. The data set represented 3778 students taking an eleventh grade test with 260 questions. The data are a matrix $X \in \{0,1\}^{3778 \times 260}$, where $X_{ij} = 1$ if student $i$ correctly answered question $j$, and is 0 otherwise. A rank 1 approximation to this matrix captures roughly 60% of the mean square. For this data set BCV chooses rank 14. The NMF is more natural for such data and we discuss this example further in Section 8.

**8. NMF examples.** We use the CLASSIC3 corpus from <ftp://ftp.cs.cornell.edu/pub/smart/> to build an NMF model where we know what the true best rank should be. This corpus consists of 3893 abstracts from journals in three different topic areas: aeronautical systems, informatics and medicine. The abstracts are typically 100–200 words, and there are roughly the same number of abstracts from each topic area. After removing stop words and performing Porter stemming [Porter (1980)], 4463 unique words remain in the corpus. Ignoring all lexical structure, we can represent the entire corpus as a $3893 \times 4463$ matrix of word counts $X^{\text{orig}}$, whose $ij$ elements is the number of times word $j$ appears in document $i$. This matrix is very sparse, with only approximately 1% of the entries being nonzero.

We construct a simulated data set $X$ from $X^{\text{orig}}$ as follows. First we split the original matrix into $X^{\text{orig}} = X_1^{\text{orig}} + X_2^{\text{orig}} + X_3^{\text{orig}}$, with one matrix per topic area. Then we fit a rank-1 NMF to each topic, $X_i^{\text{orig}} \doteq W_i H_i$. We combine these models, yielding $W = (W_1 \ W_2 \ W_3)$. The columns of $W$ represent 3 topics. We take $H \in [0, \infty)^{3 \times 4463}$ by minimizing $\|X^{\text{orig}} - WH\|$, representing each document by a mixture of these topics.

For $i = 1, \ldots, 3893$ and $j = 1, \ldots, 4463$, we sample $X_{ij}$ independently from $\text{Poi}(\mu_{ij})$ where the matrix $\mu$ equals $\varepsilon + WH$. The scalar $\varepsilon$ is the average value of the entries in $WH$, and its presence adds noise. The signal $WH$ has rank 3, but in the Poisson context, additive noise raises the rank of $\mu = \mathbb{E}(X)$ to 4.

The true error $\|\widehat{X}_k - \mu\|_F^2 / (mn)$ between $\mu$ and the $k$-term NMF approximation of $X$ is shown in Figure 7. We see that it is minimized at $k = 3$. BCV curves are also shown there. With $(2 \times 2)$-fold and $(3 \times 3)$-fold BCV, the minimizer is at $k = 3$. For ($5 \times 5$)-fold BCV, the minimizer is at $k = 4$. While this is the true rank, it attains a higher MSE by about 50%. Ten independent repetitions of the process gave nearly identical results. The BIC methods are designed for a Gaussian setting, but it is interesting to apply them here anyway. All of them chose $k = 3$.

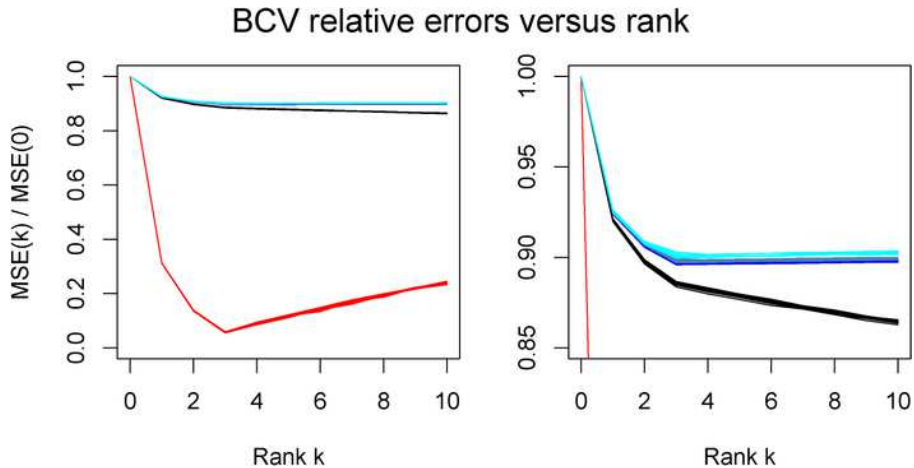

FIG. 7. *The red curves in the left panel show the true normalized squared error loss* MSE($k$)/MSE(0) *for the estimated NMFs of ranks 0 through 10 fit to the synthetic* CLASSIC3 *data. The black curves show the naive estimates of loss. The blue curves, from dark to light, show estimates based on ($5 \times 5$)-fold, ($3 \times 3$)-fold and ($2 \times 2$)-fold BCV. There were 10 independently generated instances of the problem. The right panel zooms in on the upper portion of the left panel.*

8.1. *Educational testing data.* Here we revisit the educational testing example mentioned for BCV of the SVD. We applied BCV to the NMF model using a least squares criterion and both the plain residuals (5.2), as well as the ones from (5.3) where the held out predictions are constrained to be a product of nonnegative factors.

Figure 8 shows the results. For each method we normalize the BCV squared error by $\|X\|^2$, the error for a rank 0 model. Surprisingly, the plain SVD gives a better hold out error. This suggests that factors not constrained to be nonnegative predict this data better, despite having the possibility of going out of bounds. The comparison between the two NMF methods is more subtle. Both attempt to estimate the error made by NMF on the original data. The less constrained method shows worse performance when the rank is very high. While it is a less realistic imitator of NMF, that fact may make it less prone to over-fitting.

**9. Discussion.** We have generalized the hold-out-one method of Gabriel (2002) for cross-validating the rank of truncated SVD to $r \times s$ hold-outs and then to other outer product models.

The problem of choosing the holdout size remains. Holdouts that are too small appear more prone to overfitting, while holdouts that are too large are more prone to underfitting, especially when the number of important outer product terms is not small compared to the dimensions of the held in data matrix. Until a better understanding of the optimal holdout is available, we recommend a $(2 \times 2)$-fold or $(3 \times 3)$-fold BCV, the latter coming close to the customary 10-fold CV in terms of the amount of work required. In our examples $(2 \times 2)$-BCV gave the most accurate recovery of $\mu$, except in the small SVD example.

We find that BCV works similarly to CV. In particular, the criterion tends to decrease rapidly toward a minimum as model complexity increases and then increase slowly thereafter. As a model selection method, CV works similarly to Akaike's Information Criterion (AIC) [Akaike (1974)], and differently from the Bayesian Information criterion (BIC) of Schwarz (1978). The BIC is better at finding the true model, when it is a subset of those considered, while CV and AIC are better for picking a model to minimize risk. We expect that BCV will be like CV in this regard. See Shao (1997) for a survey of model selection methods for i.i.d. sampling.

Both cross-validation and the bootstrap are sample reuse methods. When there is a symmetry in how rows and columns are sampled and interpreted, then one might choose to resample individual matrix elements or resample

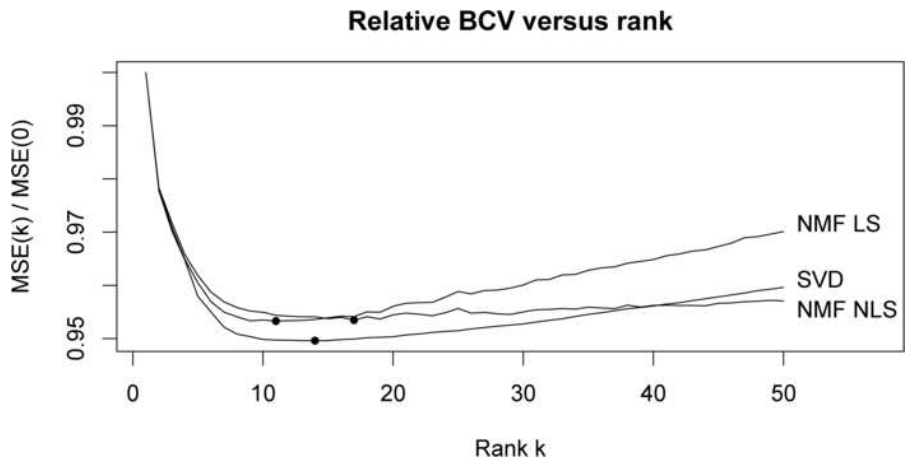

FIG. 8. *This figure plots relative mean squared error under BCV for ranks 1 to 50 on the educational testing data. The three curves are for truncated SVD, NMF with least squares residuals (5.2), and NMF with nonlinear least squares residuals (5.3). The minima, marked with solid points, are at 14, 11 and 17 respectively.*

BI-CROSS-VALIDATION OF SVD AND NMF 31

rows independently of columns. Some recent work by McCullagh (2000) has shown that in crossed random effects settings, resampling matrix elements is seriously incorrect, while resampling rows and columns independently is approximately correct. Owen (2007) generalizes those results to heteroscedastic random effects in a setting where many, even most, of the row column pairs are missing.

In the present setting it does not seem "incorrect" to leave out single matrix elements, though it may lead to over-fitting. Indeed, the method of Gabriel (2002) by leaving out a $1 \times 1$ submatrix does exactly this. Accordingly, a strategy of leaving out scattered subsets of data values, as, for example, Wold (1978) does, should also work. There are some advantages to leaving out an $r$ by $s$ submatrix though. Practically, it allows simpler algorithms to be used on the retained $(m-r) \times (n-s)$ submatrix. In particular, for the SVD there are algorithms guaranteed to find the global optimizer when blocks are held out. Theoretically, it allows one to interpret the hold out prediction errors as residuals via MacDuffee's theorem, and it allows random matrix theory to be applied.

**Acknowledgments.** We thank Alexei Onatski for sharing early results of his random matrix theory, and we thank Susan Holmes, Olga Troyanskaya and Lek-Heng Lim for pointing us to some useful references. We benefited from many helpful comments by the reviewers at AOAS, who brought the papers by Bai and Ng (2002) and Minka (2000) to our attention, and suggested looking beyond simple type I residuals. Thanks to Ed Haertel for supplying the educational testing data.

---

Department of Statistics  
Stanford University  
Stanford, California 94305  
USA  
E-mail: owen@stat.stanford.edu  
patperry@stanford.edu